# Improved scaling of the scrape-off layer particle flux width by the Bayes' theorem on EAST

D. C. Liu[1,3], X. Liu[2, *], L. Wang[2,3], and X. F. Zheng[1]

[1] *School of Physics and Electronic Information, Anhui Normal University, Wuhu 241002, People's Republic of China*

[2]*Institute of Plasma Physics, Chinese Academy of Sciences, Hefei 230031, People's Republic of China*

[3]*Institute of Energy, Hefei Comprehensive National Science Centre, Hefei 230001, People's Republic of China*

*E-mail: xliu@ipp.ac.cn

**Abstract**

The study on scaling the scrape-off layer (SOL) power width ($\lambda_q$) is crucial for deepening the understanding of the SOL particle and heat transports. Due to the sparse distribution of the divertor Langmuir probes (Div-LPs) and the erosion of probe tips during the long-pulse high-performance operations on EAST, the estimation of SOL particle flux width ($\lambda_{js}$, used to approximate $\lambda_q$) from the measured ion saturation current density profile ($j_s$) usually has relatively large uncertainty. This paper introduces a maximum a posteriori (MAP) estimation method based on the Bayes' theorem to reduce the fitting uncertainty for $\lambda_{js}$ (the fitting accuracy increases ~30% compared with the traditional ordinary least squares estimation). With the new estimation method and the FreeGS equilibrium code, the databases in [Liu X *et al*., Nucl. Fusion 64 (2024) 026002] are updated, which are further used to scale $\lambda_{js}$. Compared with the old $\lambda_{js}$ scalings for the L-mode and H-mode databases in deuterium and helium plasmas, the updated $\lambda_{js}$ scalings show better regression quality and slightly different regression results. The deuterium and helium databases can be unified to get the L-mode $\lambda_{js}^L = 0.11\bar{L}_c^{1.06}\bar{n}_e^{0.35}Z^{0.32}P_{SOL}^{0.25}\bar{p}^{-0.26}$ and H-mode $\lambda_{js}^H = 0.11\bar{L}_c^{1.28}\bar{n}_e^{0.56}Z^{0.36}P_{SOL}^{0.30}$ scalings, where $\bar{L}_c$ is the averaged SOL connection length, $\bar{n}_e$ is the line-averaged electron density, $Z$ is the charge number, $P_{SOL}$ is the power crossing the last closed flux surface, and $\bar{p}$ is the core-averaged plasma pressure. The updated scalings reveal that: i) $\lambda_{js}$ has a strong scaling dependence on the SOL connection length suggesting the missing scaling dependence on the machine size for the Eich scaling; ii) the helium $\lambda_{js}$ is slightly larger than the deuterium $\lambda_{js}$. Further discussions on extrapolation of the unified scalings to ITER give $\lambda_q$ ~ 6 and 13 mm for ITER L-mode (plasma current $I_p$ = 12 MA) and H-mode ($I_p$ = 15 MA) baseline scenarios. Furthermore, the scalings for integrated particle flux width are also given in this paper.

# 1. Introduction

In future magnetically confined fusion devices, it will be challenging to control the heat load on divertor targets. To solve this problem, it is necessary to deepen the understanding of the physics for particle and heat transports in the boundary plasmas, especially in the scrape-off layer (SOL). The studies on scaling the SOL heat flux width ($\lambda_q$) experimentally [1-13] are crucial for understanding the underlying physics of SOL particle and heat transports. $\lambda_q$ is typically measured by diagnostics, such as infra-red (IR) camera [1-7,10,11], divertor Langmuir probes (Div-LPs) [7-9,11-13], fast thermocouples [9], and reciprocating Langmuir probes (Rec-LPs) on the low-field side [7,8]. In EAST's 2019 experiment campaign, it is challenging to obtain heat flux profile due to the following facts: i) the IR camera has the problem from the heterogeneous light reflected from the upper tungsten divertor; ii) the fast thermocouples are not installed; iii) the Rec-LPs are seldomly used. Consequently, the SOL particle flux width ($\lambda_{js}$) measured by the Div-LPs is used to approximate $\lambda_q$ in the latest study [12]. Despite of this, the accurate measurement of $\lambda_{js}$ by the Div-LPs embedded in EAST's upper divertor remains challenging due to the erosion of probe tip during the long-pulse high-performance operations and the sparse distribution of the Div-LPs. To solve this issue, the following methods have been developed: i) combining two sets of Div-LPs that distributed symmetrically in the toroidal direction to reduce the measurement uncertainty [13]; ii) proposing a scaling law to calibrate the effective collection area of the probe tip [14]; iii) developing a method for correcting the particle flux measurements based on adjacent probe measurements by using the Eich function (see equation (1)) in Ohmic discharges [12]. In previous studies [12,13] the ordinary least squares (OLS) estimation is used to fit the measured ion saturation current density profile ($j_s$) by the following Eich function [4,12],

$$j_s(r) = \frac{j_{s,0}}{2} \exp\left[\left(\frac{S_{js}}{2\lambda_{js}}\right)^2 - \frac{r-r_0}{\lambda_{js}}\right] \times \mathrm{erfc}\left(\frac{S_{js}}{2\lambda_{js}} - \frac{r-r_0}{S_{js}}\right) + j_{s,BG}, \qquad (1)$$

where $r \equiv R - R_{LCFS}$ is the radial distance to the last closed flux surface (LCFS) mapped to the outboard midplane (OMP), $r_0$ is the location of the fitted LCFS, $S_{js}$ is the divertor particle flux spreading factor, $j_{s,0}$ is the peak $j_s$ at the LCFS and $j_{s,BG}$ is the background $j_s$.

Due to the sparse distribution of the Div-LPs on the tungsten divertors, there are limited measurement points covering the divertor particle flux profile. Thus, the evaluation of $\lambda_{js}$ by the OLS estimation might have considerable uncertainty, especially when $\lambda_{js}$ and/or $S_{js}$ are small. Based on the scaling databases in Ref. [12], this paper utilizes the maximum a posteriori (MAP) estimation based on Bayes' theorem to reduce the evaluation uncertainty of $\lambda_{js}$ and $S_{js}$, updates the databases, and performs the scalings of divertor particle flux footprint widths with the new databases. Compared with the OLS estimation, the MAP estimation has advantage of samples with small size and data with high noise, thereby improving the reliability of the scalings. This paper is organized as follows: section 2 introduces the estimation method based on Bayes' theorem; section 3 introduces and discusses the experimental results; section 4 summarizes the whole paper.

## 2. Estimation of divertor particle flux footprint widths based on Bayes' theorem.

### 2.1 Posterior estimation

For convivence, $j_s = \{j_s^i | i = 1 \ldots N\}$ (where i=1…N corresponds to the probe channel number) is introduced to describe the $j_s$ profile, and the Eich function is defined as $f(C)$ where $C = [j_{s,0}, \lambda_{js}, S_{js}, r_0]$. For the $i$-th probe, $j_s^i = \frac{I_s^i}{A^i}$, where $I_s^i$ is the ion saturation current measured by Div-LPs, and $A^i$ is the effective collection area of the probe tip. Due to the erosion of the probe tip during one experiment campaign, the actual effective collection area $A_{real}^i$ is changing, leading to the following measurement error in $j_s^i$,

$$j_s^i = \frac{I_s^i}{A^i} = \frac{I_s^i}{A_{real}^i} \frac{A_{real}^i}{A_{real}^i + A_{error}^i} = j_{s,real}(1 + \beta^i), \tag{2}$$

$$\beta^i = -\frac{A_{error}^i}{A_{real}^i + A_{error}^i}. \tag{3}$$

According to equations (1~3), $j_{s,real}^i$ and $j_s^i$ can be written as,

$$j_{s,real}^i = f^i(C_{real}) + \varepsilon, \varepsilon \sim \mathcal{N}(0, \sigma_{Is}^2), \tag{4}$$

$$j_s^i = f^i(C_{real}) + w^i, w^i = \frac{\beta^i}{1+\beta^i} j_s^i + \varepsilon. \tag{5}$$

Here, $\varepsilon$ represents the perturbation of $I_s^i$, which assumes to follow a normal distribution. When $\beta^i = 0$, equation (5) degenerates to equation (4) and $j_s^i = j_{s,real}^i$. Based on equation (4), the probability density function (PDF) of $j_s$ can be expressed as,

$$p(j_s; C) = \prod_i \frac{1}{\sqrt{2\pi}\sigma_{Is}} e^{-\frac{1}{2}\left(\frac{j_s^i - f^i(C_{real})}{\sigma_{Is}}\right)^2}. \tag{6}$$

The maximum likelihood estimation obtained from equation (6) is equivalent to the OLS estimation, that is,

$$\hat{C}_{OLS} = \arg\min\left\{\sum_i \left(j_s^i - f^i(C_{real})\right)^2\right\}. \tag{7}$$

When $\beta^i \neq 0$, $j_s^i \neq j_{s,real}^i$, the conditions for OLS estimation are not satisfied, meaning that the estimated parameters may not be accurate. Therefore, it is necessary to analyze the perturbation term $w^i$ to improve the estimation reliability. For simplicity, assume $A_{error}^i \sim \mathcal{N}(0, \alpha^2)$. Then,

$$w^i \sim \mathcal{N}\left(0, \left(\sigma_{js}^i\right)^2\right), \tag{8}$$

$$\sigma_{js}^i = \sqrt{(\gamma j_s^i)^2 + \sigma_{Is}^2}, \tag{9}$$

$$\gamma^2 = Var\left(\frac{\beta^i}{1+\beta^i}\right) = \frac{\alpha^2}{A_{real}^2}. \tag{10}$$

The variance of the perturbation term $w^i$, denoted as $(\sigma_{js}^i)^2$, consists of two components: the variance due to the collection area error $(\gamma j_s^i)^2$ and the variance due to the plasma perturbations $(\sigma_{Is}^2)$.

Given equations (8~10) and the regression model in equation (5), according to Bayes' theorem and assuming that the estimated parameters $\boldsymbol{C} \sim \mathcal{N}(u^j, (\sigma_C^j)^2)$ for $j = 1\ldots K$, the posterior PDF of $\boldsymbol{C}$ is given by,

$$p(\boldsymbol{C}|\boldsymbol{j_s}) \sim p(\boldsymbol{j_s}|\boldsymbol{C})p(\boldsymbol{C}) = \prod_i \frac{1}{\sqrt{2\pi}\sigma_{js}^i} e^{-\frac{1}{2}\left(\frac{j_s^i - f^i(\boldsymbol{C})}{\sigma_{js}^i}\right)^2} \times \prod_j \frac{1}{\sqrt{2\pi}\sigma_C^j} e^{-\frac{1}{2}\left(\frac{C^j - u^j}{\sigma_C^j}\right)^2}. \tag{11}$$

With the above equation, one can choose either the minimum mean square error (MMSE) estimation or the MAP estimation to evaluate $\boldsymbol{C}$. However, solving for MMSE requires multiple integrals of the PDF, which is typically feasible only using numerical methods and may result in poor accuracy. Therefore, the simpler MAP estimation is utilized, which seeks the estimation parameters that maximize the PDF in equation (11), that is,

$$\widehat{\boldsymbol{C}}_{MAP} = arg\,min\left\{\sum_i \left(\frac{j_s^i - f^i(\boldsymbol{C})}{\sigma_{js}^i}\right)^2 + \sum_j \left(\frac{C^j - u^j}{\sigma_C^j}\right)^2\right\}. \tag{12}$$

Compared with the OLS estimation given by equation (7), the above equation incorporates the weight of the observational data and adds a regularization term (prior term) on top of the OLS estimation.

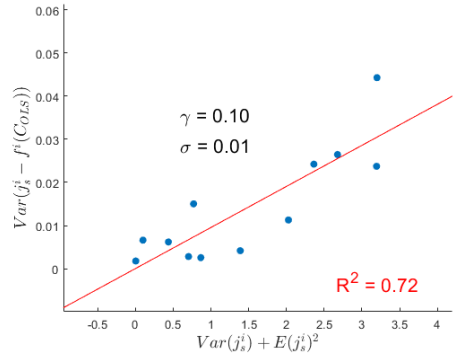

Figure 1. The linear regression result of $Var\left(j_s^i - f^i(\boldsymbol{C_{OLS}})\right)$ against $Var\left(j_s^i\right) + E\left(j_s^i\right)^2$ for dataset $D_{Ohmic}$ extracted from Ref. [12].

The MAP estimation requires the knowledge of $\gamma$, $\sigma_{Is}$ and the prior information of $\boldsymbol{C}$. The prior information of $\boldsymbol{C}$ can be approximated using the corresponding data from the scaling databases in Ref. [12]. The acquisition of $\gamma$ and $\sigma_{Is}$ is more complex. Assuming that $j_s^i$ is a random variable, the variance of equation (5) yields,

$$Var\left(j_s^i - f^i(\boldsymbol{C_{real}})\right) = \gamma^2 \left(Var(j_s^i) + E(j_s^i)^2\right) + \sigma_{Is}^2. \tag{13}$$

Then, $\gamma$ and $\sigma_{Is}$ can be obtained through linear regression by analyzing a large number of $j_s$ profiles. The unknown $\boldsymbol{C_{real}}$ is approximated by $\boldsymbol{C_{OLS}}$. Note that this approximation introduces some errors into $\gamma$ and $\sigma_{Is}$. Figure 1 shows the linear regression results using the data in Ohmic discharges (defined as dataset $D_{Ohmic}$). It is seen that a relatively good linear relationship is observed, supporting that the assumptions in the MAP estimation are reasonable.

Another method to obtain $\gamma$ and $\sigma_{Is}$ is by comparing a simulated dataset $D_{sim}$ with the experimental dataset $D_{Ohmic}$. The process for generating the dataset $D_{sim}$ is as follows: i) randomly generate $\boldsymbol{C_{real}}$ from a Gaussian distribution; ii) sample $f^i(\boldsymbol{C_{real}})$ ($i = 1\sim12$) from $f(\boldsymbol{C_{real}})$ based on the probe location; iii) add perturbations to $f^i(\boldsymbol{C_{real}})$ by using the predefined $\gamma$ and $\sigma_{Is}$ to obtain $j_s^i$ based on equation (5). With the generated dataset $D_{sim}$, $\boldsymbol{C_{OLS}}$ can be obtained by the OLS estimation. The predefined values of $\gamma$ and $\sigma_{Is}$ can be obtained, if the variances $Var\left(j_s^i - f^i(\boldsymbol{C_{OLS}})\right)$ between datasets $D_{sim}$ and $D_{Ohmic}$ are close for all probes. Table 1 lists the parameters used in the MAP estimation in this paper.

Table 1. The values of the parameters used in the MAP estimation

| $\gamma$ | $\sigma_{Is}$ | $j_{s,0}[Acm^{-2}]$ | $\lambda_{js}$ [mm] | $S_{js}$ [mm] | $r_0$ [mm] |
|---|---|---|---|---|---|
| 0.10 | 0.03 | 7.9±2.94 | 12.96±4.08 | 3.20±1.87 | -2.69±1.43 |

## 2.2 Performance comparisons between OLS and MAP estimations

With the dataset $D_{sim}$, the performance and applicability of the MAP estimation can be evaluated. Figure 2 shows the distributions of $\lambda_{js}$ and $S_{js}$ obtained using OLS and MAP estimations. It is seen that the MAP estimation has better performance than the OLS estimation in terms of both mean absolute error (MAE) and correlation coefficient (Corr) when compared with the real distribution. In figure 2(b), the distribution of $S_{js}$ obtained from the OLS estimation has an anomalous peak (similar to the $S_{js}$ distribution in figure 5(b) of Ref. [12]). This is due to the overfitting caused by the sparse distribution of the probes.

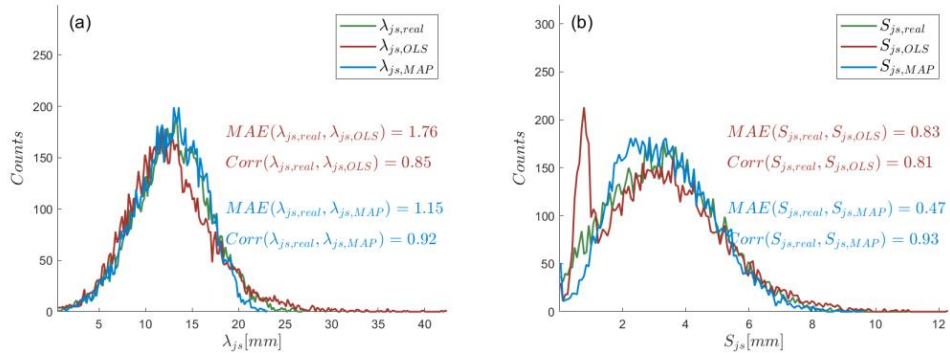

Figure 2. The distributions and errors of (a) $\lambda_{js}$ and (b) $S_{js}$ obtained from the OLS and MAP estimations based on the generated dataset $D_{sim}$.

The higher accuracy of the MAP estimation compared with the OLS estimation mainly results from two factors: i) the MAP estimation accounts for the variances of the probe collection area error and of the measurements due to plasma perturbations; ii) the MAP estimation incorporates the prior information of $C$, which constrains the range of the estimation. In general, the closer the estimated parameters to the prior information, the better the estimation accuracy. Figure 3(a) compares the MAE of the MAP estimation with that of the OLS estimation as $\lambda_{js}$ varies based on additionally generated datasets on needs. It is evident that for $\lambda_{js} \in [0.72, 25.2]$ mm, the error of the MAP estimation is lower than that of the OLS estimation. According to the parameter distributions of $\lambda_{js}$ in figures 5 and 6 of Ref. [12], the values of $\lambda_{js}$ for the EAST discharges being used in this paper are generally covered within this range.

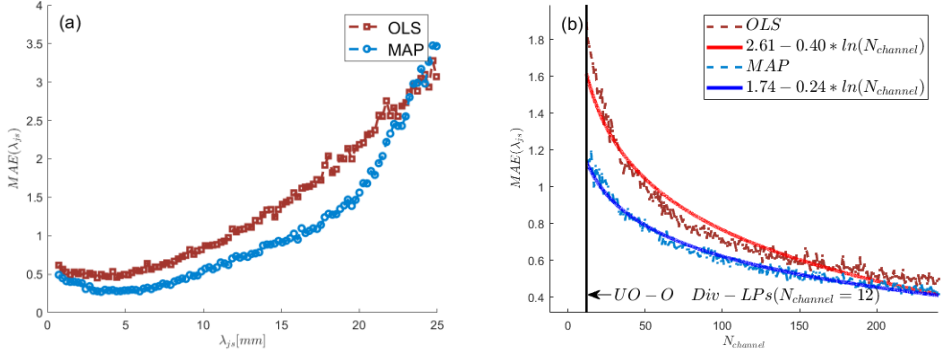

Figure 3. (a)The MAE of OLS and MAP estimations with respect to $\lambda_{js}$. (b)The MAE of OLS and MAP estimations with respect to $N_{channel}$. The vertical black line indicates the actual $N_{channel}$ for the Div-LPs being used in this paper.

It is well known that the performance of OLS and MAP estimations improves with the increase of the number of samples ($N_{channel}$). Specifically, the weight of the observed data grows while the weight of the prior terms decreases as $N_{channel}$ increases for the MAP estimation. When $N_{channel}$ reaches to a certain level, the influence of the prior terms becomes negligible, and the MAP estimation effectively becomes a classical estimation method. Regarding of this, the performance of the MAP estimation should be comparable to that of the OLS estimation. Figure 3(b) compares the MAE of $\lambda_{js}$ for the OLS and MAP estimations with respect to $N_{channel}$. As $N_{channel}$ increases, the MAE values of $\lambda_{js}$ for both methods decrease and finally saturate. The decreasing trends approximately follow the logarithmic functions (see the solid lines in figure 3(b)). The OLS estimation saturates more quickly with approximately 1.6 times faster than that of the MAP estimation. It is also seen that the MAP estimation requires smaller $N_{channel}$ value compared with the OLS estimation for similar MAE value. More specifically, the value of $N_{channel}$ should be 40 for the OLS estimation, whose estimation accuracy can be comparable to that for the MAP estimation and the MAP estimation has approximately 30% improvement in accuracy compared with the OLS estimation for the Div-LPs being used in this paper ($N_{channel} = 12$).

## 3. Experimental results and discussions

### 3.1 Comparison of SOL power width scalings by using OLS and MAP estimations

This paper replicates the data selection process in Ref. [12]: i) for each valid deuterium and helium discharges with ion $B \times \nabla B$ drifting upward in the EAST 2019 experiment campaign, several $j_s$ profiles are uniformly selected on the flat-top stage to fit $\lambda_{js}$ and the discharge is initially selected based on the fitting quality; ii) each initially selected discharge is further classified by confinement type; iii) for each confinement type in each selected discharge, $j_s$ profiles are averaged and fitted with an interval of 50 ms and a secondary selection process is performed based on the fitting quality. The detailed experimental conditions and the data selection process can be found in Ref. [12]. In this subsection, only the OLS estimation is replaced by the MAP estimation, and one criterion in the secondary selection is changed from $R^2 > 0.9$ to $R^2 > 0.85$ (the MAP estimation typically has a lower coefficient of determination $R^2$ compared with that for the OLS estimation), which removes approximately 7% data compared with the old databases constructed in Ref. [12]. In the new databases, the values of parameters such as $\lambda_{js}$ and $S_{js}$ are replaced. However, the other parameters, like the toroidal magnetic field ($B_t$), the poloidal magnetic field ($B_p$) at the LCFS on the OMP, the plasma current ($I_p$), the safety factor at 95% poloidal flux ($q_{95}$), the line-averaged electron density ($\bar{n}_e$), the plasma stored energy ($W_{MHD}$), the power crossing the LCFS ($P_{SOL}$), the divertor leg length ($L_{div,leg}$, the distance from the X-point to the outer strike point), the plasma elongation ($\kappa$), and the plasma triangularity ($\delta_{top}$) *et al*. remain unchanged.

The similar backward non-linear regressions in Ref. [12] are repeated for the $\lambda_{js}$ scalings (in deuterium L-mode (D-L), deuterium H-mode (D-H), helium L-mode (He-L) and helium H-mode (He-H) plasmas) with the new databases and the results are listed in table 2. Compared with the old scalings by the OLS estimation, the scaling parameters for the new scalings by the MAP estimation change slightly, but the regression reliability increases, especially for the D-L and He-H scalings. According to the univariate non-linear regression results for the D-L and He-H databases in figure 4, it is found that the most significantly improved scaling parameter (leads to the improvement of the scaling reliability for $\lambda_{js}$) for the MAP estimation is $\bar{n}_e$ and $L_{div,leg}$ for the D-L and He-H databases, respectively. Except for $L_{div,leg}$, the scaling dependences of $\lambda_{js}$ on the other parameters are weakened, especially for $\bar{n}_e$ and $q_{95}$ in He-L plasmas and $W_{MHD}$ in deuterium plasmas. According to the univariate regression results, the scaling on $\bar{n}_e$ and $q_{95}$ in the He-L plasmas as well as the scaling on $W_{MHD}$ in the D-L plasmas show a slight decrease in $R^2$. This leads to a reduction in the dependence of the scalings, but their absences have limited impact on the scalings since these parameters have a relatively small effect in the corresponding scalings. For example, removing $\bar{n}_e$ and $q_{95}$ from the He-L scaling results in the decrease of $R^2$ only about 0.1. Overall, the $\lambda_{js}$ scalings obtained using the MAP estimation show better quality.

Table 2. The scaling results of $\lambda_{js}$ obtained using the OLS (Ref. [12]) and MAP estimations.

| Type | Method | Scalings | $R^2$ |
|---|---|---|---|
| D-L | OLS | $\lambda_{js}^{D-L}[mm] = (13.77 \pm 1.56)\bar{n}_e^{0.59\pm0.02} W_{MHD}^{-0.47\pm0.02} L_{div,leg}^{1.27\pm0.06}$ | 0.53 |
| D-L | MAP | $\lambda_{js}^{D-L}[mm] = (21.64 \pm 1.77)\bar{n}_e^{0.53\pm0.01} W_{MHD}^{-0.31\pm0.01} L_{div,leg}^{1.24\pm0.04}$ | 0.64 |
| D-H | OLS | $\lambda_{js}^{D-H}[mm] = (1.31 \pm 0.06)\bar{n}_e^{0.75\pm0.01} W_{MHD}^{-0.54\pm0.03} P_{SOL}^{0.48\pm0.01}$ | 0.78 |
| D-H | MAP | $\lambda_{js}^{D-H}[mm] = (1.78 \pm 0.07)\bar{n}_e^{0.67\pm0.01} W_{MHD}^{-0.48\pm0.02} P_{SOL}^{0.43\pm0.01}$ | 0.84 |
| He-L | OLS | $\lambda_{js}^{He-L}[mm] = (1.71 \pm 0.11)\bar{n}_e^{0.41\pm0.02} W_{MHD}^{-0.25\pm0.01} L_{div,leg}^{0.75\pm0.02} P_{SOL}^{0.19\pm0.01} q_{95}^{1.25\pm0.03}$ | 0.70 |
| He-L | MAP | $\lambda_{js}^{He-L}[mm] = (8.33 \pm 0.42)\bar{n}_e^{0.26\pm0.01} W_{MHD}^{-0.20\pm0.01} L_{div,leg}^{0.95\pm0.01} P_{SOL}^{0.20\pm0.01} q_{95}^{0.65\pm0.02}$ | 0.79 |
| He-H | OLS | $\lambda_{js}^{He-H}[mm] = (28.42 \pm 0.85)\bar{n}_e^{0.41\pm0.02} L_{div,leg}^{0.79\pm0.03}$ | 0.43 |
| He-H | MAP | $\lambda_{js}^{He-H}[mm] = (16.88 \pm 1.24)\bar{n}_e^{0.43\pm0.02} W_{MHD}^{-0.43\pm0.03} L_{div,leg}^{0.99\pm0.02}$ | 0.71 |

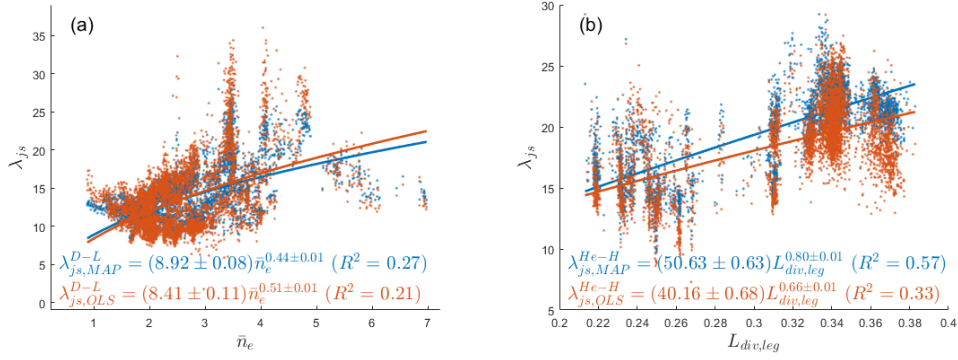

Figure 4. The univariate regression results of $\lambda_{js}$ against (a) $\bar{n}_e$ in D-L plasmas and (b) $L_{div,leg}$ in He-H plasmas using the OLS and MAP estimations.

## 3.2 Scaling of the SOL footprint widths with extended databases

Theoretically, $\lambda_q$ or $\lambda_{js}$ is closely related to the SOL connection length ($L_c$). Although $L_{div,leg}$ and $q_{95}$ in table 2 are indirect proxies for $L_c$, $L_c$ was not calculated and used as a scaling parameter in the previous study. This paper calculates $L_c$ using the FreeGS code [23] by averaging the traced field lines starting from the OMP to the outer-upper divertor. To calculate $L_c$, the original EFIT equilibrium is used as the input, and the poloidal beta ($\beta_p$), $I_p$ and the location of the LCFS are set as constraints remaining unchanged. Since the FreeGS code utilizes free boundary condition, the new equilibrium differs from the old EFIT equilibrium that are used for evaluating some of the parameters ($\lambda_{js}$, $S_{js}$, $q_{95}$, $W_{MHD}$ and $L_{div,leg}$ et al.) in the old scaling databases. In this sense, these parameters in the old scaling databases are recalculated with the new equilibrium reproduced by the FreeGS code. Table 3 compares the scaling results for $\lambda_{js}$ evaluated by the MAP estimation between the old EFIT (picked from table 2) and the new FreeGS equilibria. The obtained scalings do not change much, meaning that the change of the equilibrium from EFIT to FreeGS does not influence the $\lambda_{js}$ scalings significantly.

Table 3. Comparison of the results for $\lambda_{js}$ scalings using the MAP estimation between the EFIT and FreeGS equilibria.

| Type | Equilibrium | Scalings | $R^2$ |
|---|---|---|---|
| D-L | EFIT | $\lambda_{js}^{D-L}[mm] = (21.64 \pm 1.77)\bar{n}_e^{0.53\pm0.01} W_{MHD}^{-0.31\pm0.01} L_{div,leg}^{1.24\pm0.04}$ | 0.64 |
| | FreeGS | $\lambda_{js}^{D-L}[mm] = (30.14 \pm 2.47)\bar{n}_e^{0.57\pm0.01} W_{MHD}^{-0.32\pm0.01} L_{div,leg}^{1.36\pm0.05}$ | 0.64 |
| D-H | EFIT | $\lambda_{js}^{D-H}[mm] = (1.78 \pm 0.07)\bar{n}_e^{0.67\pm0.01} W_{MHD}^{-0.48\pm0.02} P_{SOL}^{0.43\pm0.01}$ | 0.84 |
| | FreeGS | $\lambda_{js}^{D-H}[mm] = (2.53 \pm 0.06)\bar{n}_e^{0.68\pm0.01} W_{MHD}^{-0.46\pm0.02} P_{SOL}^{0.45\pm0.01}$ | 0.85 |
| He-L | EFIT | $\lambda_{js}^{He-L}[mm] = (8.33 \pm 0.42)\bar{n}_e^{0.26\pm0.01} W_{MHD}^{-0.20\pm0.01} L_{div,leg}^{0.95\pm0.01} P_{SOL}^{0.20\pm0.01} q_{95}^{0.65\pm0.02}$ | 0.79 |
| | FreeGS | $\lambda_{js}^{He-L}[mm] = (11.57 \pm 0.50)\bar{n}_e^{0.29\pm0.01} W_{MHD}^{-0.22\pm0.01} L_{div,leg}^{0.95\pm0.01} P_{SOL}^{0.19\pm0.01} q_{95}^{0.54\pm0.02}$ | 0.79 |
| He-H | EFIT | $\lambda_{js}^{He-H}[mm] = (16.88 \pm 1.24)\bar{n}_e^{0.43\pm0.02} W_{MHD}^{-0.43\pm0.03} L_{div,leg}^{0.99\pm0.02}$ | 0.71 |
| | FreeGS | $\lambda_{js}^{He-H}[mm] = (23.51 \pm 1.15)\bar{n}_e^{0.46\pm0.02} W_{MHD}^{-0.38\pm0.03} L_{div,leg}^{1.01\pm0.02}$ | 0.71 |

Benefiting from the FreeGS code, the averaged connection length $\bar{L}_c$ (the average of $L_c$, whose radial distances to the LCFS on the OMP is from 0 to 30 mm with a 1.5 mm spacing), the core-averaged plasma pressure $\bar{p}$, the area of the LCFS $S_{LCFS} = 4\pi R_0 a\sqrt{(1+\kappa^2)/2}$, and the fraction of Greenwald density $f_{GW} = \bar{n}_e \pi a^2 I_p^{-1}$ are added to extend the new databases. Since $\bar{L}_c$ is added to replace $L_{div,leg}$ and $\bar{p}$ is highly correlated with $W_{MHD}$ ($\bar{p} \sim W_{HMD}/V$, where $V$ is the plasma volume), $L_{div,leg}$ and $W_{MHD}$ are excluded from the initial scaling parameters in Ref [12] (the other parameters are $q_{95}$, $\bar{n}_e$, $\delta_{top}$, $B_t$ and $P_{SOL}$). With the same criteria using for the backward non-linear regressions, the updated $\lambda_{js}$ scalings in D-L, D-H, He-L and He-H plasmas are obtained and the results are shown in figure 5. Compared with the FreeGS scalings in table 3, $P_{SOL}$ enters the D-L scaling, $q_{95}$ and $\bar{p}$ (replacement of $W_{MHD}$) disappear in the He-L scaling and $W_{MHD}$ is replaced by $P_{SOL}$ in the He-H scaling, leading to different values of $R^2$ for these scalings (the scaling qualities for helium scalings decrease a little bit). The updated D-H scaling shows no significant difference from the FreeGS scaling. The main differences between the updated scalings and the old scalings in Ref. [12] (see the OLS scalings in table 2) are: i) $\bar{p}$ totally vanishes in the helium scalings (a weak scaling dependence on $W_{MHD}$ exists in the old He-L scaling); ii) $q_{95}$ and $L_{div,leg}$ are combined as $\bar{L}_c$ as expected; iii) the scaling dependences on $\bar{n}_e$ are weaker, especially for the helium scalings (the possible reason has been discussed in section 3.1); iv) $P_{SOL}$ becomes an inevitable scaling parameter, just like $\bar{n}_e$.

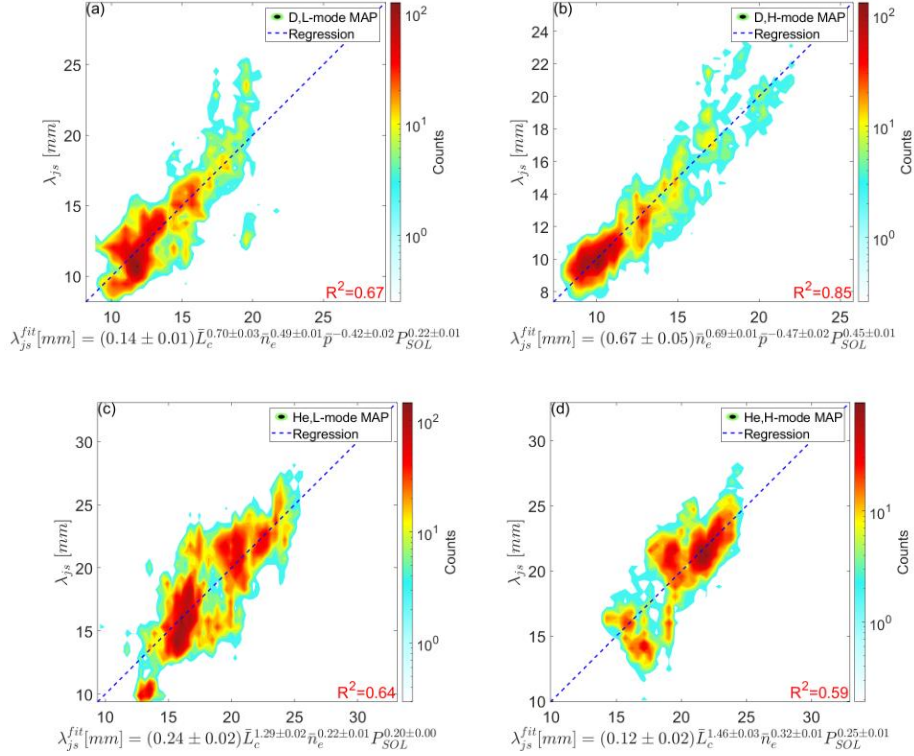

Figure 5. The scaling results of $\lambda_{js}$ for the (a) deuterium L-mode, (b) deuterium H-mode, (c) helium L-mode and (d) helium H-mode extended databases.

In Ref. [12], the deuterium and helium databases are attempted to be combined to get unified L-mode and H-mode scalings, but the key scaling parameter, charge number $Z$ (mass number doesn't influence $\lambda_{js}$ according to the discussion in Ref. [12]), is automatically removed in the backward non-linear regressions. With the extended databases, $Z$ is kept in the unified $\lambda_{js}$ scalings (see figure 6). The scaling exponent of $Z$ is very close to the theoretical and experimental results in Ref. [2], which demonstrates that helium $\lambda_{js}$ is slightly larger than deuterium $\lambda_{js}$, just like previously published experimental results [16,17]. The unified deuterium and helium $\lambda_{js}$ scalings are,

$$\lambda_{js}^L[\text{mm}] = 0.11(\bar{L}_c[\text{m}])^{1.06}(\bar{n}_e[10^{19}\text{m}^{-3}])^{0.35}Z^{0.32}(P_{SOL}[\text{MW}])^{0.25}(\bar{p}[\text{MPa}])^{-0.26}, \quad (14)$$

for deuterium and helium L-mode databases ($R^2 = 0.72$) and

$$\lambda_{js}^H[\text{mm}] = 0.11(\bar{L}_c[\text{m}])^{1.28}(\bar{n}_e[10^{19}\text{m}^{-3}])^{0.56}Z^{0.36}(P_{SOL}[\text{MW}])^{0.30}, \quad (15)$$

for deuterium and helium H-mode databases ($R^2 = 0.84$). Apparently, we see the nearly proportional scaling dependence of $\lambda_{js}$ on $\bar{L}_c$, which has been previously reported on TCV [17] and confirms the hypothesis in the HESEL simulations [18,19] that the well-known negative scaling dependence of $\lambda_q$ on $B_p$ or $I_p$ (or positive scaling dependence on the safety factor) is actually the scaling dependence on the SOL connection length. This further supports that there is a missing positive scaling dependence of $\lambda_q$ on the machine size in the Eich scaling [4], which likely results in an underestimated prediction of $\lambda_q$ for ITER. Additionally, it favors the turbulence-dominant explanation [19] instead of the drift-dominant explanation [20] of why simulation codes [19-21] can recover the Eich scaling for current devices but predict a much larger $\lambda_q$ for ITER. The scaling exponents for $\bar{n}_e$ and $P_{SOL}$ are also consistent with many previously published results [2,5,12,19].

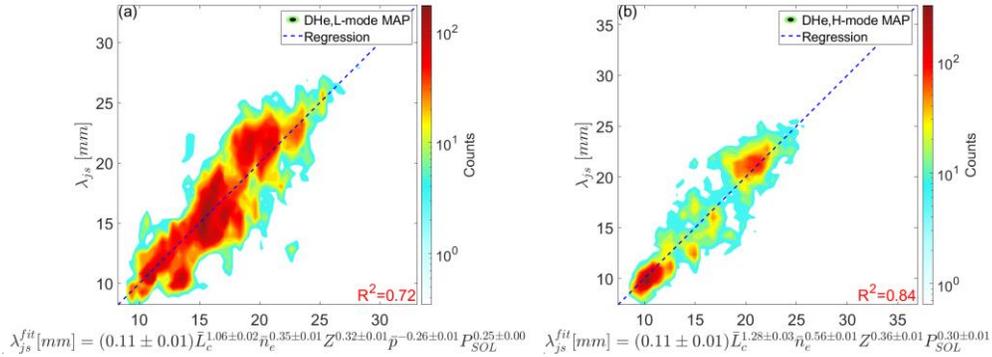

Figure 6. The results of the unified $\lambda_{js}$ scalings in the deuterium and helium (a) L-mode and (b) H-mode plasmas for the extended databases.

For better extrapolation to the other devices, especially for ITER, dimensionless parameters can be used to replace the plasma parameters in the unified scalings (equations (14) and (15)), i. e., $\bar{n}_e$ and $\bar{p}$ can be replaced by $f_{GW}$ and $\beta_p$, respectively. Additionally, $P_{SOL}/S_{LCFS}$ is used to replace $P_{SOL}$. Compared with the scaling results in figure 6, the scaling dependence on $\bar{L}_c$ is slightly enhanced, which results in the inclusion of $1/I_p$ in $f_{GW}$. The replacement of $\bar{n}_e$ to $f_{GW}$ and the replacement of $P_{SOL}$ to $P_{SOL}/S_{LCFS}$ do not violate the scaling dependences much. But the replacement of $\bar{p}$ to $\beta_p$ shows significantly different results: $\beta_p$ appears in the H-mode scaling and the scaling dependence on $\beta_p$ becomes stronger for the L-mode scaling. According to Ref. [19], $B_t$ shall be considered for extrapolation to ITER but the variation of $B_t$ is small in the databases. Take $B_t^{-0.3}$ in the HESEL scaling. The updated scalings are,

$$\lambda_{js}^L[\text{mm}] = 1.05(B_t[\text{T}])^{-0.3}(\bar{L}_c[\text{m}])^{1.20}f_{GW}^{0.34}\beta_p^{-0.30}Z^{0.28}(P_{SOL}/S_{LCFS}[\text{MW}\cdot\text{m}^{-2}])^{0.26}, \quad (16)$$

for deuterium and helium L-mode databases ($R^2 = 0.73$) and,

$$\lambda_{js}^H[\text{mm}] = 0.90(B_t[T])^{-0.3}(\bar{L}_c[m])^{1.31}\beta_p^{-0.68}f_{GW}^{0.55}Z^{0.29}(P_{SOL}/S_{LCFS}[\text{MW}\cdot m^{-2}])^{0.25}, \quad (17)$$

for deuterium and helium H-mode databases ($R^2 = 0.85$). Extrapolation of ITER L-mode $\lambda_q$ for 12MA baseline scenario [11] by equation (16) gives $\lambda_{js,ITER}^L \approx 17.51$ mm. Similarly, equation (17) gives $\lambda_{js,ITER}^H \approx 39.67$ mm for ITER 15MA baseline scenario ($Q$=10) [4,22]. Since EAST $\lambda_{js}$ is normally ~3 times larger than the published $\lambda_q$ scalings [12], the prediction of ITER L-mode and H-mode $\lambda_q$ are $\lambda_{q,ITER}^L \sim 6$ mm and $\lambda_{q,ITER}^H \sim 13$ mm. The L-mode prediction is two times larger than that for the multi-machine L-mode scaling [11] but the H-mode prediction is very close to the simulation results [19,20].

As mentioned previously, the MAP estimation has better performance than the OLS estimation when $\lambda_{js}$ and/or $S_{js}$ are small. This motivates us to perform similar scalings for $S_{js}$. However, $S_{js}$ is found to only dependent on $L_{div,leg}$ but has very low correlation with $\bar{L}_c$. Since $S_{js}$ is about 3 times smaller than $\lambda_{js}$, this might suggest that the performance of MAP estimation is intrinsically limited by the hardware characteristics (sparse distribution of Div-LPs). Instead, the integrated particle flux width $\lambda_{js,int}$ can be used to include the information of $S_{js}$ in the unified scalings and the results are listed in table 4. Compared with the $\lambda_{js}$ scalings, the presences of the scaling parameters for $\lambda_{js,int}$ remain unchanged, but the scaling dependencies of nearly all parameters are weakened (especially for the H-mode scalings) and the value of $R^2$ slightly decreases. Additionally, in the $\lambda_{js,int}$ scalings for H-mode, the scaling dependence on $Z$ increases. These changes somehow reflect the poor scaling reliability for $S_{js}$.

Table 4. The unified scaling results for $\lambda_{js,int}$.

| Type | Parameters | Scalings | $R^2$ |
|---|---|---|---|
| L-mode | dimensional | $\lambda_{js,int}^L[\text{mm}] = (0.30 \pm 0.02)\bar{L}_c^{0.86\pm0.02}\bar{n}_e^{0.27\pm0.01}\bar{p}^{-0.27\pm0.01}Z^{0.28\pm0.01}P_{SOL}^{0.22\pm0.00}$ | 0.64 |
| L-mode | dimensionless | $\lambda_{js,int}^L[\text{mm}] = (1.66 \pm 0.10)\bar{L}_c^{1.05\pm0.02}f_{GW}^{0.27\pm0.01}\beta_p^{-0.33\pm0.02}Z^{0.23\pm0.01}(P_{SOL}/S_{LCFS})^{0.24\pm0.01}$ | 0.64 |
| H-mode | dimensional | $\lambda_{js,int}^H[\text{mm}] = (0.94 \pm 0.08)\bar{L}_c^{0.80\pm0.03}\bar{n}_e^{0.36\pm0.01}P_{SOL}^{0.17\pm0.01}Z^{0.42\pm0.01}$ | 0.75 |
| H-mode | dimensionless | $\lambda_{js,int}^H[\text{mm}] = (2.92 \pm 0.30)\bar{L}_c^{0.81\pm0.03}f_{GW}^{0.33\pm0.01}\beta_p^{-0.41\pm0.02}Z^{0.37\pm0.01}(P_{SOL}/S_{LCFS})^{0.14\pm0.01}$ | 0.75 |

## 4. Summaries

In this paper, the evaluation uncertainty of $\lambda_{js}$ and $S_{js}$ from the $j_s$ profiles measured by the sparsely distributed UO-O Div-LPs on EAST is reduced by the Bayes' theorem. The formulas to perform the MAP estimation to obtain $\lambda_{js}$, $S_{js}$ and the related fitting parameters are derived. The parameters (see table 1) used in the MAP estimation are determined by comparing the Ohmic databases extracted from Ref. [12] with the generated databases. It is demonstrated that the MAP estimation has higher fitting accuracy compared with the OLS estimation, which results from the inclusion of the information of the probe collection area error and the plasma perturbations and the inclusion of the prior information to constrain the estimation range. For $\lambda_{js} \in [0.72, 25.2]$ mm (covering the $\lambda_{js}$ values used in this paper) and the employed UO-O Div-LPs (the number of probes $N_{channel} = 12$), the evaluation accuracy of the MAP estimation is approximately 30% higher than that of the OLS estimation.

The databases in Ref. [12] are updated using the MAP estimation method and the comparison results of the $\lambda_{js}$ scalings with those in Ref. [12] (the OLS estimation was employed) show better regression quality (the regression parameters do not change). To include the SOL connection length in the $\lambda_{js}$ scalings, The FreeGS code is employed with the EFIT equilibrium as input, while the plasma current, the poloidal beta, and the LCFS are set as constraints and remain unchanged. The change of equilibrium to FreeGS code alters the calculation of some of the regression parameters, but the regression results of $\lambda_{js}$ do not change significantly. Benefiting from the FreeGS code, $L_{div,leg}$ and $W_{MHD}$ are replaced by more commonly used scaling parameters, that are $\bar{L}_c$ and $\bar{p}$ respectively. The backward non-linear regression results for the D-L ($R^2 = 0.67$), D-H ($R^2 = 0.85$), He-L ($R^2 = 0.64$) and He-H ($R^2 = 0.59$) plasmas are,

$$\begin{aligned}
\lambda_{js}^{D-L}[mm] &= 0.14(\bar{L}_c[m])^{0.70}(\bar{n}_e[10^{19}m^{-3}])^{0.49}(\bar{p}[MPa])^{-0.42}(P_{SOL}[MW])^{0.22}, \\
\lambda_{js}^{D-H}[mm] &= 0.67(\bar{n}_e[10^{19}m^{-3}])^{0.69}(\bar{p}[MPa])^{-0.47}(P_{SOL}[MW])^{0.45}, \\
\lambda_{js}^{He-L}[mm] &= 0.24(\bar{L}_c[m])^{1.29}(\bar{n}_e[10^{19}m^{-3}])^{0.22}(P_{SOL}[MW])^{0.20}, \\
\lambda_{js}^{He-L}[mm] &= 0.12(\bar{L}_c[m])^{1.46}(\bar{n}_e[10^{19}m^{-3}])^{0.32}(P_{SOL}[MW])^{0.25}.
\end{aligned} \quad (18)$$

The main differences of the updated scalings compared with the previous study in Ref. [12] are: i) $q_{95}$ and $L_{div,leg}$ are combined as $\bar{L}_c$ as expected; ii) $P_{SOL}$ becomes an inevitable scaling parameter; iii) the dependence of $\lambda_{js}$ on plasma stored energy are gone for the helium databases. Further investigation shows that the L-mode ($R^2 = 0.72$) and H-mode ($R^2 = 0.84$) databases can be unified with the deuterium and helium plasmas and the results are,

$$\begin{aligned}
\lambda_{js}^{L}[mm] &= 0.11(\bar{L}_c[m])^{1.06}(\bar{n}_e[10^{19}m^{-3}])^{0.35}Z^{0.32}(P_{SOL}[MW])^{0.25}(\bar{p}[MPa])^{-0.26}, \\
\lambda_{js}^{H}[mm] &= 0.11(\bar{L}_c[m])^{1.28}(\bar{n}_e[10^{19}m^{-3}])^{0.56}Z^{0.36}(P_{SOL}[MW])^{0.30}.
\end{aligned} \quad (19)$$

The scaling exponent of $Z$ is very close to the theoretical and experimental results, which demonstrates that helium $\lambda_{js}$ is slightly larger than deuterium $\lambda_{js}$. The nearly proportional dependence of $\lambda_{js}$ on $\bar{L}_c$ supports the hypothesis that the negative scaling dependence of $\lambda_q$ on $B_p$ or $I_p$ is actually the dependence on the SOL connection length [18,19], which further suggests that there might be a missing scaling parameter that related to the machine size in the Eich scaling (leading to the underestimation of ITER $\lambda_q$). This result also supports the turbulence-dominant explanation [19] for why simulation codes can recover the Eich scaling on current devices but predict a much larger $\lambda_q$ for ITER. The scaling exponents of $\bar{n}_e$ and $P_{SOL}$ are also consistent with the previously published results [2,5,12,19]. To extrapolate the scalings to ITER, the dimensionless parameters $f_{GW}$ and $\beta_p$ are replaced by $\bar{n}_e$ and $\bar{p}$, respectively. Combined with the previous studies, the dimensionless scalings predict $\lambda_q \sim$ 6 mm and 13 mm for ITER L-mode [11] and H-mode [22] baseline scenarios. The L-mode prediction is about twice as large as the multi-machine L-mode scaling in Ref. [11], while the H-mode prediction is very close to simulation results [19,20]. The attempt to scale $S_{js}$ fails, which might be related to the intrinsic nature of the relatively poor distribution of the UO-O Div-LPs. Instead, the scalings for the integrated particle flux width are as follows,

$$\begin{aligned}\lambda^L_{js,int}[\text{mm}] &= 0.30(\bar{L}_c[m])^{0.86}(\bar{n}_e[10^{19}\text{m}^{-3}])^{0.27}Z^{0.28}(P_{SOL}[\text{MW}])^{0.22}(\bar{p}[\text{MPa}])^{-0.27}, \\ \lambda^H_{js,int}[\text{mm}] &= 0.94(\bar{L}_c[m])^{0.80}(\bar{n}_e[10^{19}\text{m}^{-3}])^{0.36}Z^{0.42}(P_{SOL}[\text{MW}])^{0.17}.\end{aligned} \quad (20)$$

It is seen that the scaling dependences of nearly all parameters are weakened compared with equations (19), and helium $\lambda_{js,int}$ is also slightly larger than deuterium $\lambda_{js,int}$.

## Acknowledgments

This work was supported by the National Natural Science Foundation of China (Nos. 12275312, 12005260, 12474255). This work was also partially supported by the Institute of Energy, Hefei Comprehensive National Science Center (Anhui Energy Laboratory) under Grant No. 21KZS202 and the HFIPS Director's Fund with Grant No. BJPY2023A05.

## References

1  Eich T *et al.*, Phys. Rev. Lett. 107 (2011) 215001.

2  Fundamenski W *et al.*, Nucl. Fusion 51 (2011) 083028.

3  Makowski M.A *et al.*, Phys. Plasmas 19 (2012) 056122.

4  Eich T *et al.*, Nucl. Fusion 53 (2013) 093031.

5  Eich T *et al.*, Journal of Nuclear Materials (2013) S72–S77.

6  Scarabosio A *et al.*, J. Nucl. Mater. 438 (2013) S426–30.

7  Wang L *et al.*, Nucl. Fusion 54 (2014) 114002.

8  Liu J B *et al.*, Fusion Eng. Des. 100 (2015) 301–6.

9  Brunner D *et al.*, Nucl. Fusion 58 (2018) 094002.

10  Silvagni D *et al.*, Plasma Phys. Control. Fusion 62 (2020) 045015.

11  J. Horacek *et al.*, Nucl. Fusion 60 (2020) 066016.

12  Liu X *et al.*, Nucl. Fusion 64 (2024) 026002.

13  Liu X *et al.*, Plasma Phys. Control. Fusion 61 (2019) 045001.

14  Meng L Y *et al.*, Nuclear Materials and Energy 27 (2021) 100996.

15  Jie HUANG *et al.*, Plasma Sci. Technol 23 (2021) 084001.

16  Sieglin B *et al.*, Plasma Phys. Control. Fusion 58 (2016) 055015.

17  Faitsch M *et al.*, Plasma Phys. Control. Fusion 60 (2018) 045010.

18  Liu X *et al.*, Phys. Plasmas 26 (2019) 042509.

19  Liu X *et al.*, Nucl. Fusion 62 (2022) 076022.

20  Ze-Yu Li *et al.*, Nucl. Fusion 59 (2019) 046014.

21. Chang C.S *et al.*, Nucl. Fusion 57 (2017) 116023.
22. A. C. C. Sips *et al.*, Phys. Plasmas 22 (2015) 021804.
23. Ben Dudson et al., FreeGS, GitHub, https://github.com/freegs-plasma/freegs